\documentstyle[aps,prl,epsf,floats]{revtex}
\begin{document}
\draft
%%
%Definitions
%%
\def\ef{\epsilon_{\rm F}}
\def\efb{\epsilon_F^{>}}
\def\rof{\rho_0(\epsilon_{\rm F})}
\twocolumn[\hsize\textwidth\columnwidth\hsize\csname@twocolumnfalse\endcsname

\title{Coherence scale of the Kondo lattice}
\author{S. Burdin${}^{1,2}$, A. Georges${}^{3}$
and D. R. Grempel${}^{1,4}$}
\address{$^1$D\'epartement de Recherche Fondamentale sur la Mati\`ere
Condens\'ee,\\ SPSMS,  CEA-Grenoble,
38054 Grenoble Cedex 9,
France.\\
${}^{2}$ Institut Laue-Langevin, B.P. 156, 38042 Grenoble Cedex 9, France.\\
${}^{3}$CNRS - Laboratoire de Physique Th{\'e}orique, Ecole Normale
Sup{\'e}rieure, 24 Rue Lhomond 75005 Paris, France.\\${}^{4}$ Service
de Physique de l'Etat Condens\'e, CEA-Saclay, 91191
Gif-sur-Yvette Cedex, France.}
\date{\today}
\maketitle
\widetext
\begin{abstract}
\noindent
It is shown that the large-N approach yields two energy scales for
the Kondo lattice model. The single-impurity Kondo temperature,
$T_K$, signals the onset of local singlet formation, while Fermi
liquid coherence sets in only below a lower scale, $T^{\star}$. At low
conduction electron density $n_c$ (``exhaustion'' limit)
, the ratio $T^{\star}/T_K$ is much
smaller than unity, and is shown to depend only on $n_c$ and not
on the Kondo coupling. The physical meaning of these two scales is
demonstrated by computing several quantities as a function of
$n_c$ and temperature.
\newline
\end{abstract}
%\pacs{71.27.+a, 71.10.Fd, 71.20.Eh}
]
\narrowtext 

How coherent quasi-particles form in the Kondo lattice has been a
long-standing issue. For a single impurity,
there is a single scale $T_K$ below which the local moment is screened
and a local Fermi liquid picture applies.
$T_K$ can be defined e.g. as the scale at
which the effective Kondo coupling becomes large. All physical
quantities (e.g. specific heat, susceptibility) obey scaling
properties as a function of $T/T_K$ in the limit where both $T$
and $T_K$ are smaller than the high-energy cutoff (bandwidth,
denoted $2D$ in the following). In
contrast, for a Kondo lattice, one may suspect that the physics is
no longer governed by a single scale. Indeed, while magnetic
moments can still be screened {\it locally} for temperatures lower
than the single-impurity Kondo scale $T_K$, the formation of a
Fermi-liquid regime (with coherent quasi-particles and a ``large''
Fermi surface encompassing both conduction electrons and the
localized spins) is a global phenomenon requiring coherence over
the whole lattice. If at all possible, it could be
associated with a much lower coherence temperature $T^{\star}$. The
situation is reminiscent of strong-coupling superconductivity in
which local pair formation may occur at a much higher scale than
$T_c$, the scale at which long-range order
sets in.

As was originally pointed out by Nozi\`eres \cite{pn1}, this issue
becomes especially relevant when few conduction electrons are
available to screen the local spins, i.e. in the limit of low
concentration ($n_c\ll 1$). In this ``exhaustion'' regime, two
possibilities arise: i) either magnetic ordering wins
over Kondo screening or ii) a paramagnetic Fermi liquid state
still manages to form, but with a much suppressed coherence scale
$T^{\star}\ll T_K$. Nozi\`eres has suggested in Ref.\cite{pn2} that
$T^{\star}\sim n_c D$ for strong coupling $J_K/D\gg 1$ (where $T_K\sim
J_K$), while $T^{\star}\sim T_K^2/D$ for weak coupling $J_K\ll D$.
Recently, several studies \cite{PAMinfd1,PAMinfd2,tz1,tz3,krish99,nrg} have
addressed this issue using dynamical mean-field theory (DMFT)
\cite{dmft}. The conclusion was that  two scales are indeed
present in the Kondo lattice, with the coherence scale with
$T^{\star}\ll T_K$ in the
``exhaustion'' regime. Because the DMFT equations require a
numerical treatment, no detailed analytical insight into these two
scales was obtained, even though the validity of the estimate \cite{pn2}
$T^{\star}\propto T_K^2/D$ was questioned
\cite{nrg}.

In this letter, we solve this problem using the slave-boson
approach, in the form of a controlled large-N solution of the
$SU(N)$ Kondo lattice model. This approach has been extensively
used in the past twenty years
\cite{largen,newns,riceueda,millis,assa,piers}. Surprisingly, 
the issue of the coherence scale and the temperature dependence of 
physical quantities has not been discussed in detail in the exhaustion
limit $n_c\ll 1$
(See, however, Ref.\ \onlinecite{claudine}). We find that
the large-N approach provides a remarkably simple and physically
transparent realization of the two-scale scenario described above.
Furthermore, because of its simplicity, it allows for an
analytical calculation of the coherence scale, which is found to
disagree (for weak coupling) with Nozi\`eres' estimate in
Ref.\cite{pn2} (while it agrees with it at strong coupling). We also
calculate the temperature dependence of several physical
quantities and find remarkable agreement with the more
sophisticated (and demanding) DMFT approach.

We consider the Kondo lattice model (KLM):
\begin{equation}
H = \sum_{k\sigma} \epsilon_k c^+_{k\sigma}c_{k\sigma} +
{{J_K}\over{N}}\sum_{i\sigma\sigma'} S_i^{\sigma\sigma'}
c^+_{i\sigma'}c_{i\sigma}
\label{ham}
\end{equation}
In this expression, the spin symmetry has been extended to $SU(N)$
($\sigma=1,\cdots,N$) and the local spins will be considered in the
fermionic representation:
$S_i^{\sigma\sigma'} =
f^+_{i\sigma}f_{i\sigma'}-\delta_{\sigma\sigma'}/2$, with the
constraint $\sum_{\sigma}f^+_{i\sigma}f_{i\sigma}=N/2$.
Standard methods \cite{largen} are used to solve this model
at large $N$: a boson field $B_i(\tau)$ (conjugate to the amplitude
$\sum_\sigma f^+_{i\sigma}c_{i\sigma}$) is introduced in order to decouple the
Kondo interaction, and the constraint is implemented through a Lagrange
multiplier field $\lambda_i(\tau)$.
For $N=\infty$, a saddle point is found at which
the Bose field condenses $\langle B_i(\tau)\rangle = r$
and the Lagrange multiplier takes a uniform,
static value $\langle i\lambda_i(\tau)\rangle=\lambda$. Two quasiparticle
bands $\omega_k^{\pm}$ are formed, solutions of
$(\omega+\lambda)(\omega + \mu -\epsilon_k)-r^2=0$. Changing variables to
$\omega_k^{\pm}$, the three saddle-point equations can be cast
in the compact form:
\begin{eqnarray}
\label{saddle}
\nonumber
\left\{-{1\over J_K},{1\over 2}, {n_c\over 2}\right\}
=&&\int_{-\infty}^{+\infty} d\omega n_F(\omega)
\rho_0(\omega + \mu -{{r^2}\over{\omega+\lambda}})\times\\
&&\left\{
{{1}\over{\omega+\lambda}}, {{r^2}\over{(\omega+\lambda)^2}},
1 \right\}.
\label{speqs}
\end{eqnarray}
Here, $n_F$ is the Fermi function, $n_c/2$ is the
conduction electron density per spin colour, and
$\rho_0(\epsilon)\equiv\sum_k\delta(\epsilon-\epsilon_k)$ is the
non-interacting density of states.

In the large-N approach, the onset of Kondo screening is signaled
by a phase transition at a critical temperature $T_K$, below which
$r(T)$ becomes non-zero. The equation for  $T_K$ is $2/ J_K = \int
d\epsilon \rho_0(\epsilon) \tanh
[(\epsilon-\mu_0)/2T_K]/(\epsilon-\mu_0)$, with $\mu_0$ the
non-interacting chemical potential (at $T=T_K$). This equation
coincides with that for the single-impurity case: Kondo screening of
individual local moments starts taking place in this approach {\it
precisely at the single-impurity Kondo scale}. We have derived an
explicit expression for $T_K$ in the weak-coupling regime $J_K\ll
D$:
\begin{eqnarray}
\label{Tk} T_K&=& D e^{-1/J_K\rof}\,\sqrt{1-(\ef/D)^2} F_K(n_c),\\
\label{FK} F_K(n_c)&=&\exp{\left(\int_{-(D+\ef)}^{D-\ef}
{{d\omega}\over{|\omega|}} \frac{
\rho_0(\ef+\omega)-\rho_0(\ef)}{2 \rho_0(\ef)}\right)},
\end{eqnarray}
where $\ef$ is the non-interacting Fermi level
(given by $n_c/2 =
\int_{-D}^{\ef} d\epsilon\rho_0(\epsilon)$). This expression is
valid for an even d.o.s $\rho_0(-\epsilon)=\rho_0(\epsilon)$ which
vanishes outside the interval $-D<\epsilon <+D$. The factor $F_K$,
equal to unity for a constant density of states, varies smoothly
with $n_c$ (or $\epsilon_F$). In contrast, the prefactor $[1 -
(\ef/D)^2]^{1/2}$ vanishes in the low-density limit $n_c\to 0$,
and suppresses $T_K$.

We now consider the low-temperature limit $T\ll T_K$, in which
the large-N approach leads to an extremely simple Fermi liquid
picture \cite{largen} . It is somewhat oversimplified in that the
conduction electron self-energy
$\Sigma_c(k,\omega)=r^2/(\omega+\lambda)$
 is purely real (finite lifetime effects
are absent at $N=\infty$) and momentum independent. Even so, it
captures some of the most important features of the problem. The
zero-frequency shift
$\Sigma_c(\omega=0,T=0)=r(T=0)^2/\lambda(T=0)$ is precisely such
that the Fermi surface has a {\it large volume} containing both
conduction electrons and local spins. Indeed, adding the last two
saddle point equations in (\ref{speqs}), one obtains:
$\mu-\Sigma_c(\omega=0,T=0) = \efb$ where $\efb$ is the
non-interacting value of the Fermi-level corresponding to
$(n_c+1)/2$ fermions per spin color. As detailed below, all
physical quantities at $T=0$ are directly related to a single
energy scale, proportional to the boson condensation
amplitude $T^{\star}= r^2(T=0)/D$. It is possible to derive an
analytical expression for $T^{\star}$ in the weak coupling
limit, which is valid for a general (bounded) density of states
and arbitrary density $n_c$. This expression, which has apparently
not been reported before, reads:
\begin{eqnarray}
\label{tstar} T^{\star}&=& D  e^{-1/J_K\rof}\, \left(1 +
\ef/D\right)\frac{\Delta\ef}{D} F^{\star}(n_c),\\ \label{Fstar}
F^{\star}(n_c)&=& \exp{\left(\int_{-(D+\ef)}^{\Delta\ef}
{{d\omega}\over{|\omega|}} \frac{
\rho_0(\ef+\omega)-\rho_0(\ef)}{\rho_0(\ef)}\right)},
\end{eqnarray}
where $\Delta\ef = \ef^> - \ef$. As $F_K$, $F^{\star}$ varies smoothly with 
$n_c$ \cite{grilli}.
The total density of states at
the Fermi level $\rho(\omega=0)=\rho_{cc}(0)+\rho_{ff}(0)$ is
given by: $\rho(0)=\rho_0(\efb)/Z_c$, with $Z_c$ the conduction
electron quasiparticle residue $1/Z_c\equiv
1-\partial\Sigma_c(\omega)/\partial\omega|_{\omega=0} = 1+(\efb-\ef)^2/r_0^2$. In
the weak coupling limit, all physical quantities at $T=0$ are
directly related to $\rho(0)\simeq
\rho_0(\ef^>)(\Delta\ef)^2/(T^{\star}D)$, and  are thus 
renormalized by the ratio $T^{\star}/D$. For instance, the $f$-electron
susceptibility and the specific heat coefficient (per spin
color), are given by $\chi_f\propto\rho(0)$ and
$\gamma=\pi^2\rho(0)/3$. The Drude weight may also be computed
with the result $D_R \propto T^{\star}D
\rho_0(\ef^>)/(\Delta\ef)^2$.

The physical content of these expressions and of
Eqs.(\ref{Tk}-\ref{Fstar}) is that {\it two different energy
scales} are relevant for the Kondo lattice model: one
($T_K$) is associated with the onset of local Kondo screening; the
other ($T^{\star}$) with Fermi liquid coherence and the
behaviour of physical quantities at $T=0$. These two scales have
{\it the same} exponential dependence 
on 
$J_K/D$ at weak coupling, but very different
dependence on the conduction electron density in the
``exhaustion'' limit $n_c\ll 1$, in which $T^{\star}\ll T_K$. This is in
qualitative agreement with Nozi\`eres proposal in \cite{pn2}, but
not with his estimate $T^{\star}/T_K\propto T_K/D$ (which would thus
depend on the coupling). We find that the ratio
$T^{\star}/T_K$ depends only on $n_c$ in this limit, in a manner which
reflects the behavior of the bare d.o.s $\rho_0(\epsilon)$ at the
bottom of the band. For a d.o.s vanishing as $(\epsilon +
D)^{\alpha}$, Eqs.(\ref{Tk}-\ref{Fstar}) yield $T^{\star}/T_K
\propto n_c^{1/[2(1+\alpha)]}$.

\begin{figure}[t]
\epsfxsize=3.3in
\centerline{\epsffile{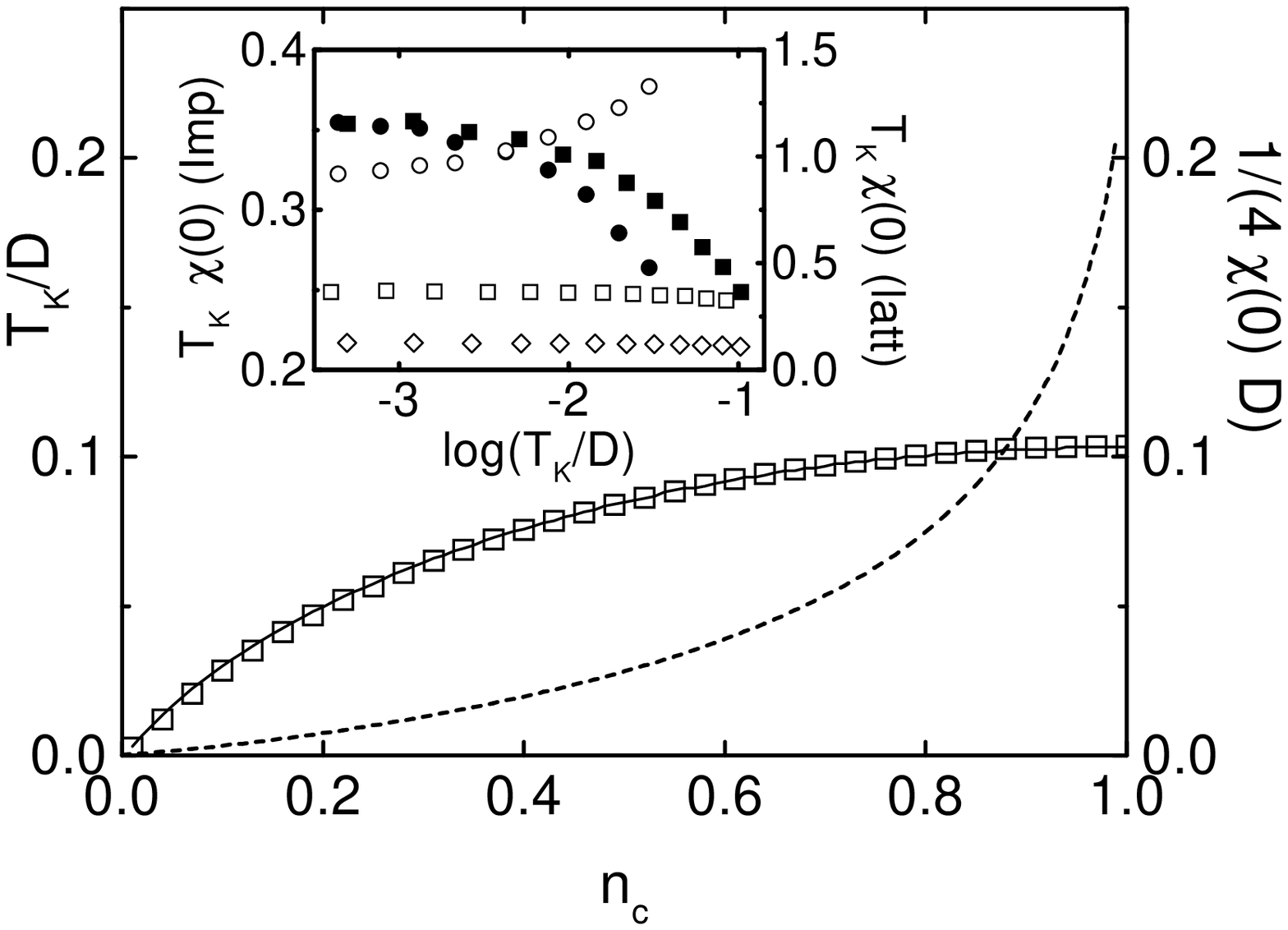}}
\caption{Solid line: density-dependence of $T_K$ (defined 
as the slave boson condensation temperature). Squares:
$1/4\chi_f(T$=0)
for a single impurity. Dashed line: $1/4\chi_f(T$=0) for the lattice. $J_K/D$=0.75.
Inset: $T_K$-dependence of $T_K\chi_f(0)$ for the impurity (left scale, solid symbols) and the lattice (right scale,
open symbols) for $n_c$=0.1, 1.0 (impurity) and
$n_c$=0.1, 0.5, 0.9 (lattice) from top to bottom.
All plots are for a semi-circular d.o.s.}
\label{fig1}
\end{figure}

\begin{figure}[t]
\epsfxsize=3in
\centerline{\epsffile{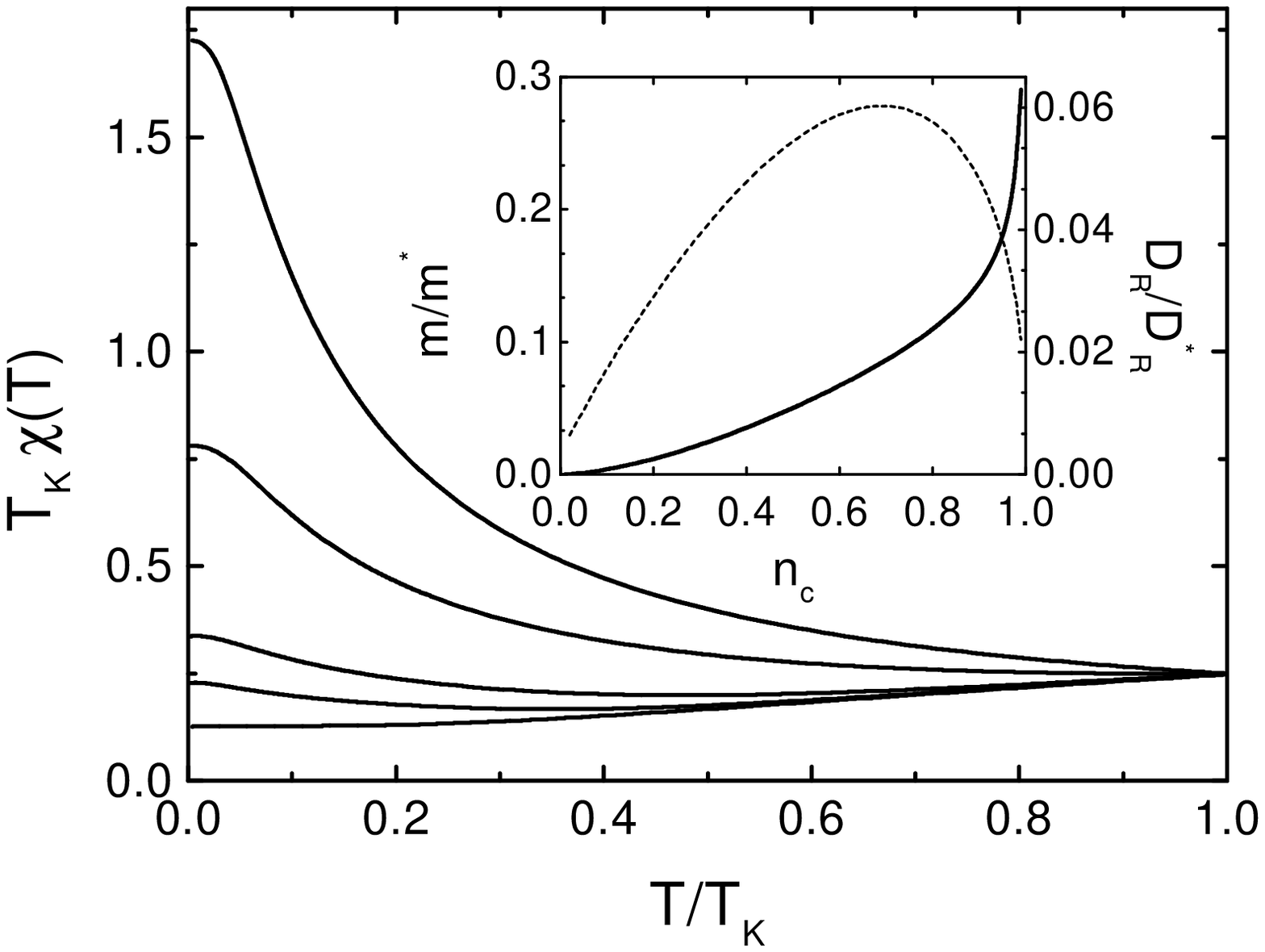}}
\caption{$T$-dependence of the $f$-electron susceptibility for
the lattice for $n_c$=0.01, 0.2, 0.5, 0.8 and 0.9, from top to
bottom. Inset: Inverse effective-mass(left scale,
full line) and Drude factor (right scale, dashed line) as functions of
density. In all plots, $J_K/D=0.75$}
\label{fig2}
\end{figure}

Fig.\,\ref{fig1} displays the $n_c$-dependence of $T_K$ and of the inverse of the $f$-electron susceptibility,
obtained by solving the large-N equations,
both for a single-impurity and for the lattice.
It is seen that 
$T_K$ and $1/\chi_f(T=0)$ are identical energy scales for the
single impurity case, but have very different density dependence
for the lattice model ($T^{\star}/T_K \propto n_c^{1/3}$ at low
density for the semi-circular d.o.s used here).
Near $n_c=1$, $T_K$ falls below $1/\chi$ 
reflecting the vanishing $\chi$ for the Kondo
insulator. These curves are remarkably similar to those obtained
by Tahvildar-Zadeh {\it et al.} in their Quantum Monte Carlo
studies of the periodic Anderson model in infinite dimensions
\cite{tz1} (see also \cite{tz1}-\cite{nrg}).
In the inset of Fig.\,\ref{fig1}, we plot the dimensionless number
$T_K\chi_f(0)$ as a function of $\ln T_K/D$. This number goes to
a universal value at weak coupling in the single-impurity case, while
it has intrinsic density dependence for the lattice. This shows 
that $T_K$ is not the appropriate low-temperature scale, especially
for $n_c\to 0$.
The inset in Fig.\,\ref{fig2} shows the effective mass ratio
$m/m^{\star}$ (computed from the specific heat) and the Drude weight,
as a function of $n_c$.
This vanishes in the limits $n_c\to 0$ and $n_c\to 1$. The effective mass $m^{\star}$ is proportional
to $e^{+1/J_K\rho_0(\epsilon_F)}$, with a density dependent
prefactor which diverges as $n_c\to 0$ and vanishes
at half filling.

\begin{figure}[t]
\epsfxsize=3.3in
\centerline{\epsffile{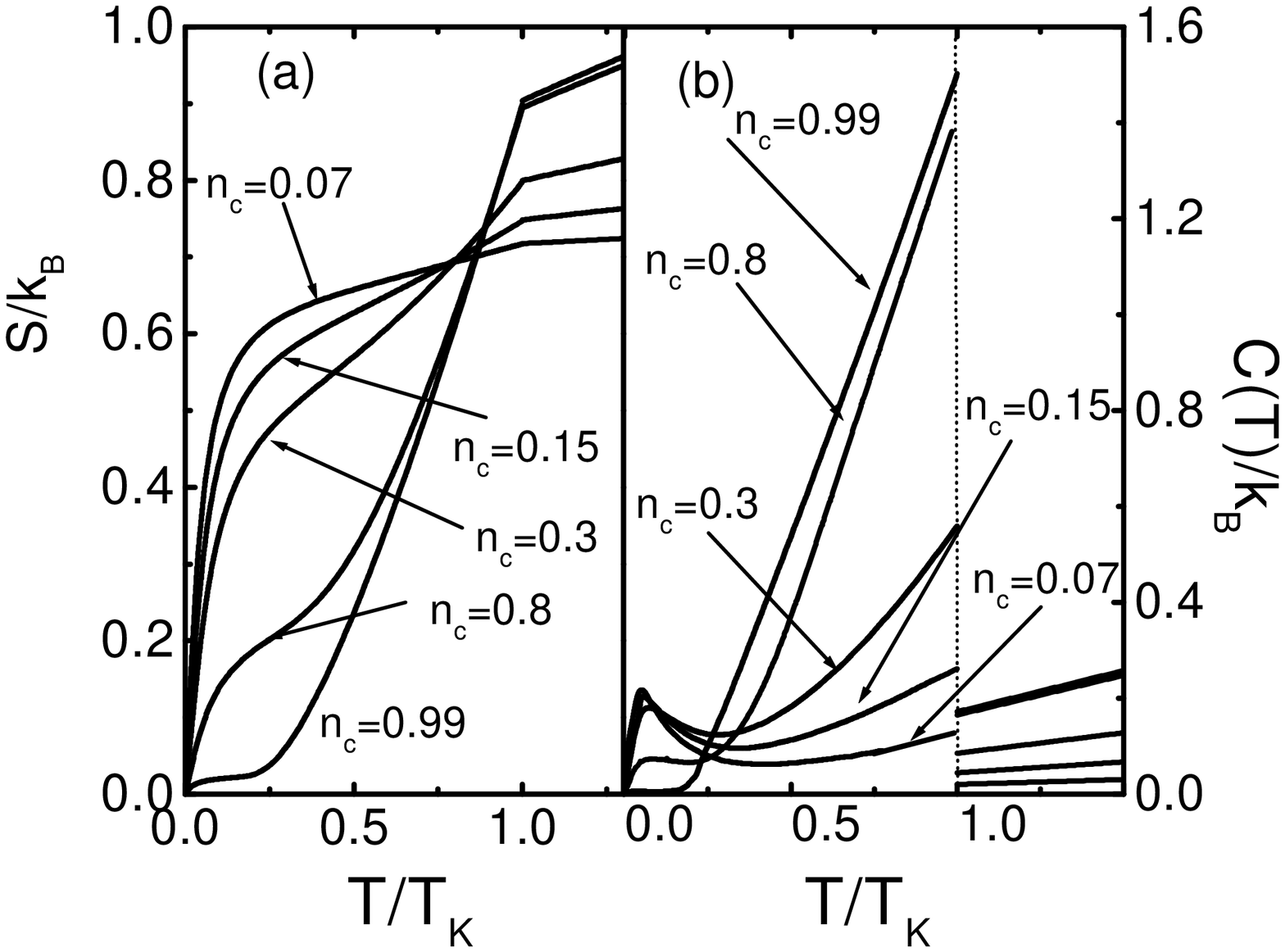}}
\caption{Entropy (a) and specific heat (b) for the lattice model for
several values of the density. $J_K/D$=0.75.}
\label{fig3}
\end{figure}

We have also studied the behavior of several physical quantities
as a function of temperature, by solving numerically
Eqs.(\ref{saddle}). Fig.\,\ref{fig2} shows the product $T_K
\chi(T)$ for the Kondo lattice, for several values of $n_c$, as a
function of $T/T_K$. All the curves merge at
$T/T_K=1$, where the boson decondenses and $\chi_f$ reaches the
free spin value ($\chi_f(T)=1/4T$ for $T>T_K$). No universal
scaling function of $T/T_K$ describes the temperature dependence
in Fig.\,\ref{fig2}, in contrast to the single-impurity case. Plotting $\chi(T)/\chi(T=0)$ does not restore scaling, since
qualitative differences in the T-dependence are seen for different
densities. Fig.\,\ref{fig3}(a) shows the $T$-dependence of the
entropy. A linear behavior $S(T)=\gamma T$ is found at
low temperature for all densities $n_c\ne 1$. The slope
$\gamma\propto \rho(0)$  decreases with increasing density as does
the temperature scale $T_F^{\star}$ up to which $S(T)$ is linear.
$T_F^{\star}$ is of the order of $T^{\star}$ ($\ll T_K$) at low $n_c$, while
it can be estimated by comparing $\gamma T$ to $e^{-T_K/T}$ 
as $T_F^{\star}\simeq T_K/|\ln(1-n_c)|\ll T_K \simeq
T^{\star}$ for $n_c\to 1$. For
$n_c\simeq 1$, $T_F^{\star}$ is a better estimate of the coherence
scale than $T^{\star}$ itself.

At low densities, after a steep initial rise, $S(T)$ remains close
to $\ln 2$ up to $T_K$. This can be interpreted in terms of the
strong-coupling picture discussed below. At higher
densities, most of the variation takes place in the range
$T_F^{\star}<T<T_K$. The specific heat $C(T)$, shown in 
Fig.\,\ref{fig3}(b), has a two-peak structure \cite{pc}: the peak at 
$T_K$ signals the onset of Kondo screening and appears in this
mean-field description as a discontinuity of $C(T)$. The second
peak, at $T_F^{\star}$, signals the Fermi liquid heavy-fermion
regime. As $n_c$ increases, weight is gradually transferred
to the high-temperature peak until the low temperature peak
disappears completely in the Kondo insulator limit.

In the strong coupling limit $J_K/D \gg 1$,
the large-N results support the physical picture proposed by
Lacroix \cite{lacroix} and further discussed in \cite{pn2}. In
this picture, the conduction electrons bind to $n_c$ spins,
forming singlets below $T\sim J_K\sim
T_K$, the binding energy of a singlet. The remaining $1-n_c$
``bachelor spins'' behave as itinerant fermions subject to a
constraint of no double occupancy. The hopping integral of the
resulting effective infinite-U Hubbard model is $t_{\rm
eff}=-t/2$. The (hole-like) sign of this matrix element implies
that these $1-n_c$ fermions have  a Fermi surface volume
corresponding precisely to $n_c + 1$ particles. In the exhaustion
limit $n_c\ll 1$, one has effectively a weakly doped $U=\infty$
Hubbard model. Solving Eqs.(\ref{speqs}) at strong-coupling yields
a quasiparticle residue $Z_c\propto n_c$, and hence a coherence
scale $T^{\star}\sim n_c D$ corresponding to a Brinkman-Rice 
estimate for this doped Mott
insulator \cite{pn2}.  Notice, however, that at finite $N$ this uniform solution may
become unstable to magnetism or phase separation of localized singlets
and unscreened spins. This picture sheds some light on the $T$-dependence of
the entropy and specific heat in the exhaustion limit reported
above. As the system goes through $T_K$, it looses the magnetic entropy ($\sim n_c\ln 2$) of the
$n_c$ bound spins. The remaining entropy ($\sim (1 - n_c)\ln2$),
is lost below $T^{\star}$. The two peaks of unequal
weight in the specific heat reflect these processes \cite{foot}.

\begin{figure}[t]
\epsfxsize=2.8in
\centerline{\epsffile{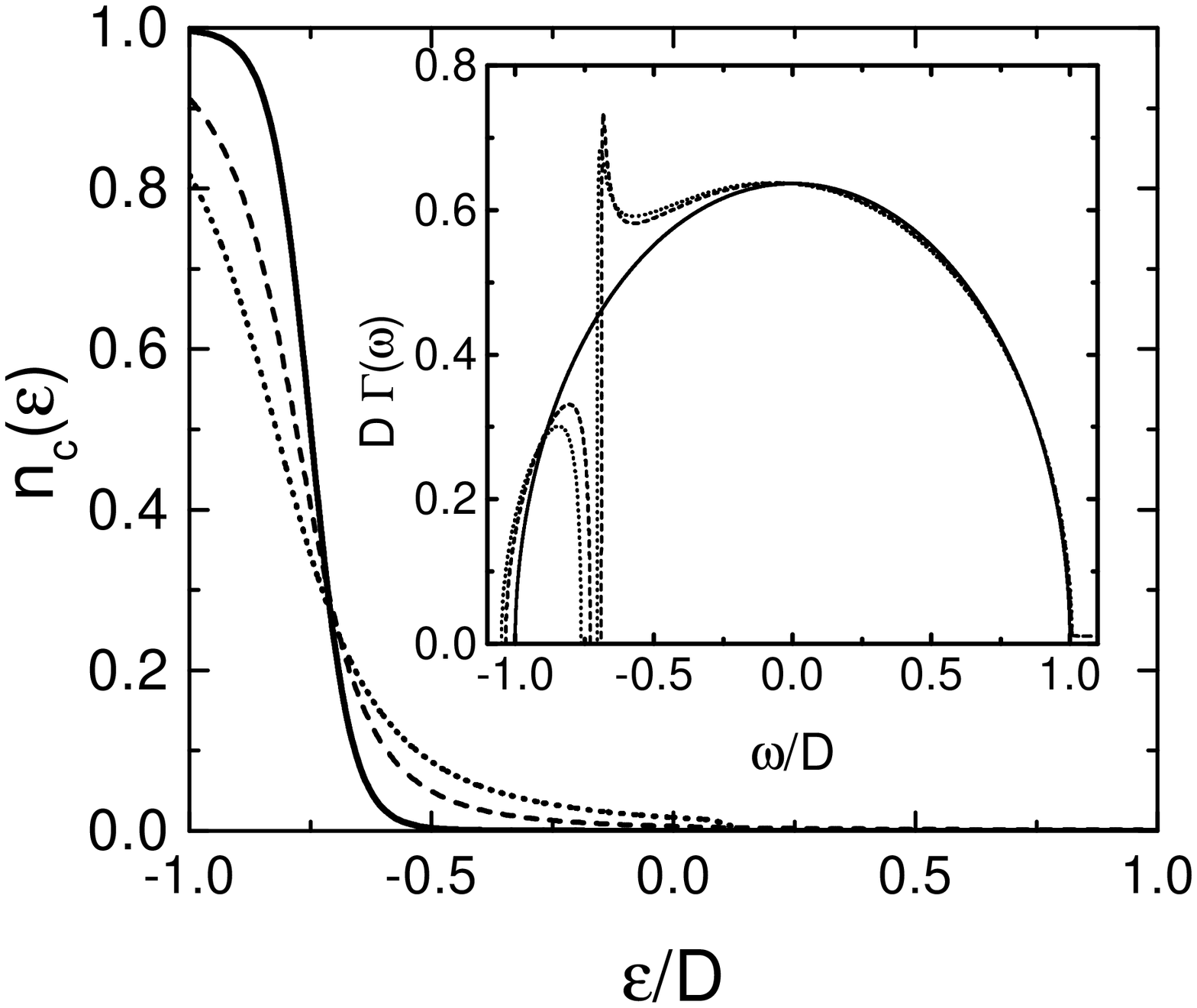}}
\caption{$n_c(\epsilon)$ for
$T/T_K$=1.0 (solid line), 0.5 (dashed line), 0.005 (dotted
line).$n_c$=0.15 Inset: Spectral density $\Gamma(\omega)$ for $T/T_K$=1 (solid
line), 0.5 (dashed line) and 0.25 (dotted line). For
$T<T_K$ there is a $\delta$-function peak in the gap not shown for
clarity. $n_c$=0.2.
In all plots $J_K/D$=0.75.}
\label{fig4}
\end{figure}

We display in Fig.\,\ref{fig4} the distribution function of
the conduction electrons $n_c(\epsilon_k)\equiv \langle c^+_k
c_k\rangle$. Very close to $T_K$, $n_c(\epsilon_k)$ has the shape
of a Fermi function centered around the {\it non-interacting}
Fermi level $\ef$, with a small thermal broadening of order $T_K$
($\ll D$ in weak coupling). As $T$ is reduced below $T_K$, weight
is transferred to scales of order $\efb$, the {\it interacting}
Fermi level associated with the large Fermi surface. The feature
at $\ef$ {\it broadens} as $T$ is decreased. When Fermi liquid
coherence establishes at $T\simeq T_F^{\star}$, a discontinuity (of
amplitude $Z_c$) develops at $\epsilon=\efb$. Finally we note
that, in the large-N approach, the Kondo lattice model can be
exactly mapped onto an effective single-impurity model coupled to
a self-consistent bath of electrons. This mapping holds more
generally for any approach in which the conduction electron
self-energy depends only on frequency, such as dynamical
mean-field theory \cite{dmft}. For a semi-circular d.o.s, the
Green's function of the self-consistent bath is ${\cal
G}_0(i\omega_n) = [i\omega_n+\mu - D^2
G_c(i\omega_n)/4]^{-1}$. The inset of Fig.\,\ref{fig4} displays 
the continuous evolution of the spectral density $\Gamma(\omega)=-\mbox{Im}{\cal
G}_0(\omega+i0^+)/\pi$ as $T$ is reduced below $T_K$. Above $T_K$,
it coincides with the non-interacting d.o.s $\rho_0$, while the
bath is split into two bands below $T_K$. The width of the lower
band depends on $n_c$ and is small for
$n_c\ll 1$, leading to the low coherence scale. The existence of a
sharp gap separating the two bands is an artifact of the large-N
limit (except at $n_c=1$). In more realistic treatments it is
replaced by a pseudo-gap \cite{tz1,krish99,nrg}.

In conclusion,  we have reconsidered the time-honored
large-N approach to the Kondo lattice model, with special emphasis
on the ``exhaustion'' limit of low electron density. We showed 
that two energy scales appear for which we have obtained explicit
analytic expressions : the Kondo scale associated
with the onset of local Kondo screening and a much lower scale
associated with Fermi liquid coherence. Physical quantities
reflect the crucial role played by these two scales. In this
approach, magnetic ordering is suppressed by quantum fluctuations 
: more elaborate treatments such as DMFT must be used
to assess whether the coherence scale can actually be reached
or magnetic ordering sets in at a higher temperature.

Useful discussions with B. Coqblin, M. Jarrell,  A. Jerez, G. Kotliar,
C. Lacroix, P. Nozi\`eres,
and A. Sengupta are gratefully acknowledged. We thank the Newton
Institute where part of this work was performed for its hospitality.

\end{document}